\documentclass[prl,showkeys,showpacs,footnoteinbib,twocolumn,unsortedaddress]{revtex4}
\usepackage{amsmath,amssymb,epic,eepic}
\usepackage{siunitx}
\usepackage[utf8]{inputenc}
\usepackage{graphicx}
\usepackage{subfigure}
\usepackage{tikz}
\usetikzlibrary{patterns}
\usetikzlibrary{positioning}
\usetikzlibrary{shapes,shapes.misc,shapes.geometric} 
\usetikzlibrary{arrows}
\usetikzlibrary{calc}
\usetikzlibrary{matrix}
\usetikzlibrary{intersections}
\usetikzlibrary{decorations,decorations.markings,decorations.pathreplacing,
decorations.pathmorphing}

\usepackage{datetime}

\newcommand{\Ge}{\mathcal{G}^\text{(e)}}
\newcommand{\DGe}{\Delta\mathcal{G}^\text{(e)}}
\newcommand{\Gee}{\mathcal{G}^\text{(2e)}}
\newcommand{\DGee}{\Delta\mathcal{G}^\text{(2e)}}
\newcommand{\ee}{\, \mathrm{e}}
\newcommand{\ii}{\mathrm{i}}
\newcommand{\dd}{\mathrm{d}}

\begin{document}

\title{Two-electron coherence and its measurement in electron quantum optics}

\author{\'{E}. Thibierge$^1$}
\author{D. Ferraro$^{1,2,3}$}
\author{B. Roussel$^{1,4}$}
\author{C. Cabart$^5$} 
\author{A. Marguerite$^5$}
\author{G. F\`eve$^5$}
\author{P. Degiovanni$^1$}

\affiliation{(1) Universit\'e de Lyon, F\'ed\'eration de Physique A.-M. Amp\`ere,\\
CNRS - Laboratoire de Physique - Ecole Normale Sup\'erieure de Lyon,\\
46 All\'ee d'Italie, 69364 Lyon Cedex 07, France}

\affiliation{(2) Aix Marseille Universit\'e, CNRS, CPT, UMR 7332, 13288 Marseille, France} 

\affiliation{(3) Universit\'e de Toulon, CNRS, CPT, UMR 7332, 83957 La Garde, France}

\affiliation{(4) Université Claude Bernard Lyon 1, \\
43 Bd du 11 Novembre 1918, 69622 Villeurbanne Cedex, France}

\affiliation{(5) Laboratoire Pierre Aigrain, Ecole Normale Sup\'erieure-PSL
Research University, CNRS, Universit\'e Pierre et Marie Curie-Sorbonne
Universit\'es, Universit\'e Paris Diderot-Sorbonne Paris Cit\'e, 24 rue
Lhomond, 75231 Paris Cedex 05, France.
}

\begin{abstract}
Engineering and studying few-electron states in ballistic conductors is a 
key step towards understanding entanglement in quantum electronic systems. In this Letter, we
introduce the intrinsic two-electron coherence of an electronic source in quantum Hall edge channels and
relate it to two-electron wavefunctions and to current noise in an Hanbury Brown--Twiss interferometer. 
Inspired by the analogy
with photon quantum optics, we propose to measure the intrinsic two-electron coherence of a source
using low-frequency current correlation measurements
at the output of a Franson interferometer. To illustrate this protocol, we discuss how it can distinguish between
a time-bin entangled pure state and a statistical mixture of time shifted electron pairs.
\end{abstract}



\maketitle

The initial development of electron quantum optics~\cite{Bocquillon:2014-1} 
has focused on
single electron coherence~\cite{Degio:2010-4,Haack:2012-2} in ballistic
quantum conductors through
Mach--Zehnder interferometry~\cite{Ji:2003-1, Roulleau:2007-2,
Roulleau:2008-2} and
electronic distribution function
measurement~\cite{Altimiras:2010-1,LeSueur:2010-1}.
With the availability
of single electron 
sources~\cite{Feve:2007-1, Blumenthal:2007-1, Hermelin:2011-1,
Hohls:2012-1, Dubois:2013-2},
Hanbury Brown--Twiss (HBT)~\cite{Bocquillon:2012-1} and 
Hong--Ou--Mandel (HOM) experiments~\cite{Bocquillon:2013-1} 
have been demonstrated.
These two-particle interference experiments are
important milestones towards single electron
tomography~\cite{Degio:2010-4}, mirroring
optical homodyne tomography~\cite{Lvovsky:2009-1}. 
Recent experimental progresses~\cite{Jullien:2014-1}
suggest that technology has reached the point
where single electron coherence can be measured using  
signal processing strategies~\cite{Ferraro:2013-1}, thus opening the way
to quantitative studies of
electronic decoherence~\cite{Wahl:2013-1,Ferraro:2014-2}
and decoherence protection techniques~\cite{Huynh:2012-1,Altimiras:2010-2}. 

The new frontier for electron quantum optics is now the
emergence of many-body physics in quantum nanoelectronics.
Beyond understanding single electron decoherence, electron quantum
optics
gives a unique and new 
insight on the build up of quantum correlations in a coherent conductor as a result of 
strong Coulomb
interactions and fermionic statistics.
For example, studying interaction induced decoherence of engineered few electron
excitations
(among which the
charge-$n$ Leviton~\cite{Keeling:2006-1,Dubois:2013-1})
is essential for understanding the interplay
of interaction effects among the 
$n$ particles and many-body effects involving all electrons
present in the Fermi sea such as decoherence an relaxation. 
By encoding all the information on $n$-electronic wave functions present in the
electronic fluid,
higher order electronic coherences~\cite{Moskalets:2014-1} defined by analogy with their
photonic counterparts~\cite{Glauber:1963-1} are the appropriate quantities to
tackle these problems.
Moreover, because entanglement criteria can be expressed in
terms of many-body correlators~\cite{Wolk:2014-1}, 
higher electronic coherences will also be instrumental in quantifying
entanglement in the electronic many-body 
fluid~\cite{Amico:2008-1}. 
From a broader perspective, understanding higher-order coherences in
presence of a Fermi sea is also relevant for quantum transport simulations with 
fermionic cold atom systems~\cite{Brantut:2012-1,Krinner:2015-1} and 
atomic HOM experiments~\cite{Lopes:2015-1}. 

\medskip

In this Letter, we make a step in this direction by discussing
the recently introduced two-electron coherence~\cite{Moskalets:2014-1}, 
an essential quantity 
to quantify two-particle
entanglement beyond the orbital point of
view~\cite{Chtchelkatchev:2002-1,Samuelsson:2003-1,
Samuelsson:2004-1,Samuelsson:2006-1,Neder:2007-2}, that is, in the time 
and energy domains. In this perspective, 
we show how to define the intrinsic two-electron coherence emitted by
an electron source and discuss how it encodes the information on
two-electron wavefunctions. We then propose to access it
through current correlations in a Franson 
interferometer~\cite{Franson:1989-1} which has also been proposed to study Cooper pair
coherence~\cite{Giovannetti:2012-1}. This naturally extends
single electron coherence measurement through average
current in Mach--Zehnder interferometry~\cite{Haack:2011-1}. 

\medskip

By analogy with quantum
optics~\cite{Glauber:1962-1,Glauber:1963-1}, two-electron coherence
is defined as~\cite{Moskalets:2014-1}:
\begin{equation}
\label{eq:definition:g2}
\Gee_\rho(1,2|1',2')=\mathrm{Tr}\left[
\psi(2)\psi(1)\rho\,\psi^\dagger(1')\psi^\dagger(2')\right]
\end{equation}
where $\psi$ and $\psi^\dagger$ are field operators for the electron fluid,
$\rho$ its density operator 
and $1=(x_1,t_1)$, $2=(x_2,t_2)$, $1'=(x'_1,t'_1)$ and $2'=(x_2',t'_2)$ 
four space-time coordinates.

\begin{figure}[htb]\centering
\small
\begin{tikzpicture}[
	scale = 0.3,
  decoration={
    markings,
    mark=at position .5 with {\arrow{stealth}}}
  ]

\begin{scope}
\draw (1.75,-5.5) node {(a) Fermi sea.} ;


\coordinate (O) at (0,0);

\coordinate (S2bottom) at (0,2);
\coordinate (S2top) at (0,3);

\coordinate (S4bottom) at (0,1.6);
\coordinate (S4midbottom) at (0,2.2);
\coordinate (S4midtop) at (0,2.8);
\coordinate (S4top) at (0,3.6);

\coordinate (F2bottom) at (0,-2.5);
\coordinate (F2top) at (0,-1.5);

\coordinate (F4top) at (0,-1.4);
\coordinate (F4midtop) at (0,-2);
\coordinate (F4midbottom) at (0,-2.6);
\coordinate (F4bottom) at (0,-3.2);

\coordinate (D1top) at (5,2);
\coordinate (D1bottom) at (5,1.5);

\coordinate (D2top) at (5,-2);
\coordinate (D2bottom) at (5,-3);


\draw [red, postaction={decorate}] (F4top) -- (D1top);
\draw (3,2) node {1};
\draw [blue, postaction={decorate}] (F4midbottom) -- (D2top);
\draw (3.5,-1.5) node {$2$};

\draw [red, dashed, postaction={decorate}] (D1bottom) -- (F4midtop) ;
\draw (4,-0.1) node {$1'$};
\draw [blue, dashed, postaction={decorate}] (D2bottom) -- (F4bottom) ;
\draw (3.5,-4) node {$2'$};

\draw [fill = gray!30, thick] (-2.5,1) rectangle (0,3);
\draw (-1.25,2) node {$(S)$};
 \fill [fill = gray!30] (-2.5,-.5) rectangle (0,-4);
\draw [thick] (-2.5,-.5) -- (0,-.5) -- (0,-4);
\draw (-1.25,-2.25) node [text width=1cm, text centered] {$(F)$};

\draw [thick, red, fill = red!20] (6,2.5)  arc (90:270:2.2 and 1.1) ;
\draw (5.2,1.4) node {$D_1$};
\draw [thick, blue, fill = blue!20] (6,-1.5)  arc (90:270:2.2 and 1.1) ;
\draw (5,-2.7) node {$D_2$};
\end{scope}

\begin{scope}[xshift=15cm]
\draw (1.75,-5.5) node {(b) Classical correlations.} ;

\coordinate (O) at (0,0);

\coordinate (S2bottom) at (0,1.75);
\coordinate (S2top) at (0,2.25);

\coordinate (S4bottom) at (0,1.6);
\coordinate (S4midbottom) at (0,2.2);
\coordinate (S4midtop) at (0,2.8);
\coordinate (S4midtop) at (0,3.6);

\coordinate (F2bottom) at (0,-2.25);
\coordinate (F2top) at (0,-1.75);

\coordinate (F4top) at (0,-2);
\coordinate (F4midtop) at (0,-2.6);
\coordinate (F4midbottom) at (0,-3.2);
\coordinate (F4bottom) at (0,-3.8);

\coordinate (D1top) at (5,2);
\coordinate (D1bottom) at (5,1.5);

\coordinate (D2top) at (5,-2.5);
\coordinate (D2bottom) at (5,-3);


\draw [red, postaction={decorate}] (S2top) -- (D1top);
\draw (D1top) ++ (-1,1) node {$1$};
\draw [blue, postaction={decorate}] (F2top) -- (D2top);
\draw (4,-1.2) node {$2$};

\draw [red, dashed, postaction={decorate}] (D1bottom) -- (S2bottom) ;
\draw (3.2,0.5) node {$1'$};
\draw [blue, dashed, postaction={decorate}] (D2bottom) -- (F2bottom) ;
\draw (3.5,-4) node {$2'$};

\draw [fill = gray!30, thick] (-2.5,1) rectangle (0,3);
\draw (-1.25,2) node {$(S)$};
 \fill [fill = gray!30] (-2.5,-.5) rectangle (0,-4);
\draw [thick] (-2.5,-.5) -- (0,-.5) -- (0,-4);
\draw (-1.25,-2.25) node [text width=1cm, text centered] {$(F)$};

\draw [thick, red, fill = red!20] (6,2.5)  arc (90:270:2.2 and 1.1) ;
\draw (5.2,1.4) node {$D_1$};
\draw [thick, blue, fill = blue!20] (6,-1.5)  arc (90:270:2.2 and 1.1) ;
\draw (5,-2.7) node {$D_2$};
\end{scope}

\begin{scope}[yshift=-11cm]
\draw (1.75,-5.5) node {(c) Quantum exchange.} ;

\coordinate (O) at (0,0);

\coordinate (S2bottom) at (0,1.5);
\coordinate (S2top) at (0,2.5);

\coordinate (S4bottom) at (0,1.6);
\coordinate (S4midbottom) at (0,2.2);
\coordinate (S4midtop) at (0,2.8);
\coordinate (S4midtop) at (0,3.6);

\coordinate (F2bottom) at (0,-2.5);
\coordinate (F2top) at (0,-1.5);

\coordinate (F4top) at (0,-2);
\coordinate (F4midtop) at (0,-2.6);
\coordinate (F4midbottom) at (0,-3.2);
\coordinate (F4bottom) at (0,-3.8);

\coordinate (D1top) at (5,2);
\coordinate (D1bottom) at (5,1.5);

\coordinate (D2top) at (5,-2);
\coordinate (D2bottom) at (5,-3);


\draw [red, postaction={decorate}] (S2top) -- (D1top);
\draw (4,3) node {1};
\draw [blue, postaction={decorate}] (F2bottom) -- (D2bottom);
\draw (D2bottom) ++ (-1,-1) node {$2$};

\draw [red, dashed, postaction={decorate}] (D1bottom) -- (F2top) ;
\draw (3,1) node {$1'$};
\draw [blue, dashed, postaction={decorate}] (D2top) -- (S2bottom) ;
\draw (D2top) ++ (-.1,1.4) node {$2'$};

\draw [fill = gray!30, thick] (-2.5,1) rectangle (0,3);
\draw (-1.25,2) node {$(S)$};
 \fill [fill = gray!30] (-2.5,-.5) rectangle (0,-4);
\draw [thick] (-2.5,-.5) -- (0,-.5) -- (0,-4);
\draw (-1.25,-2.25) node [text width=1cm, text centered] {$(F)$};

\draw [thick, red, fill = red!20] (6,2.5)  arc (90:270:2.2 and 1.1) ;
\draw (5.2,1.4) node {$D_1$};
\draw [thick, blue, fill = blue!20] (6,-1.5)  arc (90:270:2.2 and 1.1) ;
\draw (5,-2.7) node {$D_2$};
\end{scope}

\begin{scope}[xshift=15cm,yshift=-11cm]
\draw (1.75,-5.5) node {(d) Source intrinsic.} ;
%

\coordinate (O) at (0,0);

\coordinate (S2bottom) at (0,2);
\coordinate (S2top) at (0,3);

\coordinate (S4bottom) at (0,1.2);
\coordinate (S4midbottom) at (0,1.7);
\coordinate (S4midtop) at (0,2.3);
\coordinate (S4top) at (0,2.8);

\coordinate (F2bottom) at (0,-3);
\coordinate (F2top) at (0,-2);

\coordinate (F4top) at (0,-2);
\coordinate (F4midtop) at (0,-2.6);
\coordinate (F4midbottom) at (0,-3.2);
\coordinate (F4bottom) at (0,-3.8);

\coordinate (D1top) at (5,2);
\coordinate (D1bottom) at (5,1.5);

\coordinate (D2top) at (5,-2.5);
\coordinate (D2bottom) at (5,-3);


\draw [red, postaction={decorate}] (S4top) -- (D1top);
\draw (D1top) ++ (-1,1) node {$1$};
\draw [blue, postaction={decorate}] (S4midbottom) -- (D2top);
\draw (4.5,-1) node {$2$};

\draw [red, dashed, postaction={decorate}] (D1bottom) -- (S4midtop) ;
\draw (3,1) node {$1'$};
\draw [blue, dashed, postaction={decorate}] (D2bottom) -- (S4bottom) ;
\draw (3,-2.5) node {$2'$};

\draw [fill = gray!30, thick] (-2.5,1) rectangle (0,3);
\draw (-1.25,2) node {$(S)$};
 \fill [fill = gray!30] (-2.5,-.5) rectangle (0,-4);
\draw [thick] (-2.5,-.5) -- (0,-.5) -- (0,-4);
\draw (-1.25,-2.25) node [text width=1cm, text centered] {$(F)$};

\draw [thick, red, fill = red!20] (6,2.5)  arc (90:270:2.2 and 1.1) ;
\draw (5.2,1.4) node {$D_1$};
\draw [thick, blue, fill = blue!20] (6,-1.5)  arc (90:270:2.2 and 1.1) ;
\draw (5,-2.7) node {$D_2$};
\end{scope}
  
\end{tikzpicture}
\caption{\label{fig:processes} (Color online)
Contributions to the two-electron detection probability 
by detectors~$D_1$ and~$D_2$.
Full lines represent direct probability amplitudes
whereas dotted lines represent complex conjugated ones.
The electronic system is decomposed into the Fermi sea~$(F)$ 
and a source~$(S)$ emitting electron and/or hole excitations. 
The two-electron detection probability contains four types of
contributions.
}
\end{figure}
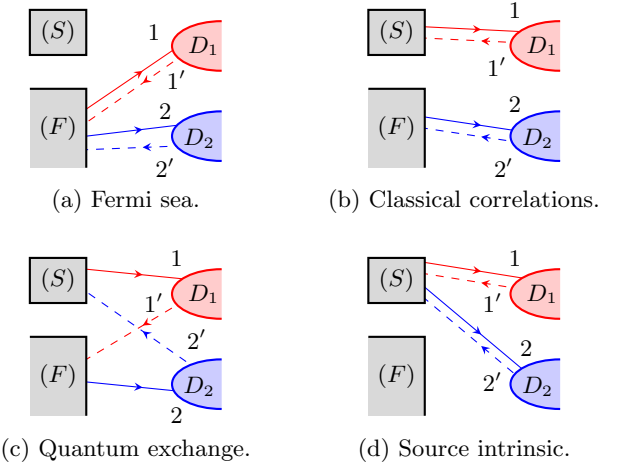

Activating a source $(S)$ generating
excitations on top of the Fermi sea alters all electronic coherences.
A source-intrinsic contribution to the total single electron coherence
$\Ge_S(1|1')=\mathrm{Tr}\left[\psi(1)\rho_S\,\psi^\dagger(1')\right]$
can be defined~\cite{Degio:2011-1,Haack:2012-2} by substracting the
Fermi sea contribution $\mathcal{G}^{(\mathrm{e})}_F$:
\begin{equation}
\DGe_S(1|1') = \Ge_S(1|1') - \Ge_F(1|1') \,.
\end{equation}
To identify the intrinsic second order coherence of the source from the
full~$\Gee_S$,
we must consider all processes
contributing to the co-detection of two electrons, depicted on Fig.~\ref{fig:processes}. 
The total two-electron coherence is the sum of
(a) the two-electron coherence of the Fermi sea;
(b) terms describing classical coincidences in which one electron 
comes from the Fermi sea and the other one from the source;
(c) quantum exchange contributions involving two-particle 
interferences between the source and the Fermi sea; 
(d) the remaining term that defines intrinsic two-electron 
coherence of the source~$\DGee_S(1,2|1',2')$:
\begin{subequations}
\label{eq:g2}
\begin{align}
\Gee_S & (1,2|1',2') = \Gee_F(1,2|1',2') 
\label{eq:g2:fermi}\\ 
+ & \,\Ge_F(1|1') \, \DGe_S(2|2') + \Ge_F(2|2') \DGe_S(1|1') \label{eq:g2:class}\\
- & \,\Ge_F(1|2') \, \DGe_S(2|1')- \Ge_F(2|1') \, \DGe_S(1|2') \label{eq:g2:exch}\\
+ & \,\DGee_S(1,2|1',2')\,.
\label{eq:g2:excess}
\end{align}
\end{subequations}
The definition of $\Delta\Gee_S$ given by Eq.~\eqref{eq:g2} is the second
in a hierarchy of definitions for the $n$-electron excess coherence with 
respect to a reference  situation, here the Fermi sea. 
All of them make sense even in the presence of interactions since their 
definition does not rely on Wick theorem.

The physical meaning of intrinsic two-electron coherence can be 
illustrated by a few examples. We consider fixed positions 
for detectors so that only time variables are mentioned. Interactions are neglected.
First, we consider an ideal single electron source emitting
a single electronic excitation in a wavepacket~$\varphi$. 
Using Wick theorem, the excess single electron coherence is found to be
$\DGe_S(t|t') = \varphi(t) \, \varphi^*(t')$
whereas $\DGee_S(t_1,t_2|t'_1,t'_2)$ vanishes as
expected from the analogy with quantum optics.
An ideal two-electron source emitting
a two-electron excitation above the Fermi sea
$\psi^\dagger[\varphi_a]\psi^\dagger[\varphi_b]|F\rangle$, 
where $\varphi_{a,b}$ are two orthogonal normalized electronic wavepackets
generates:
\begin{subequations}
\begin{align}
\label{eq:2ES:G1}
&\DGe_S(t | t') =
	\varphi_a(t) \, \varphi_a^*(t')
	+ \varphi_b(t) \, \varphi_b^*(t') \,,\\
\label{eq:2ES:G2}
&\DGee_S(t_1, t_2 | t'_1, t'_2) = \Phi_{ab}(t_1, t_2)\,\Phi_{ab}^*(t'_1, t'_2) \,;
\end{align}
\end{subequations}
where $\Phi_{ab}$ is the Slater determinant built 
from~$\varphi_a$ and~$\varphi_b$,
\begin{equation}
\label{eq:2es:wavefct}
\Phi_{ab}(t_1, t_2) = \varphi_a(t_1) \, \varphi_b(t_2)
- \varphi_b(t_1) \, \varphi_a(t_2)\,.
\end{equation}
Therefore, $\DGee_S$ gives access to the 
two-particle wave function emitted by the source, extending at the two particle level 
the relation of $\DGe_S$ to single particle wavefunctions~\cite{Degio:2010-4}. 
The case of an electron train 
built from mutually orthogonal wavepackets
extends Eq.~\eqref{eq:2ES:G2} into a sum 
involving all pairs of electronic excitations,
\begin{equation}
\label{eq:G2:train}
\DGee(t_1,t_2|t'_1,t'_2) =
	\sum_{\text{pairs }\{i,j\}} \Phi_{ij}(t_1,t_2) \,
		\Phi_{ij}^*(t'_1,t'_2)\,.
\end{equation}
The diagonal part $\DGee_S(t_1 , t_2 | t_1, t_2)$ of the intrinsic two-electron coherence encodes 
the two-particle time correlations of the electron flow.
By analogy with quantum optics, we define the degree of second order electronic coherence of the source 
by
\begin{equation}
\label{eq:def:g2}
g^{(2\text{e})}_S(t_1 , t_2) = \frac{\DGee_S(t_1 , t_2 | t_1, t_2)
	}{
	\DGe_S(t_1|t_1)\,\DGe_S(t_2|t_2)}	\,.
\end{equation}
Statistical independence of the detection events at $t_1$ and $t_2$ 
corresponds to $g^{(2\text{e})}_S(t_1,t_2)=1$
whereas antibunching is
associated with $g^{(2\text{e})}_S(t_1,t_2)<1$ and bunching with $g^{(2\text{e})}_S(t_1,t_2)>1$.

\begin{figure}[ht]
\centering
\includegraphics[scale=1]{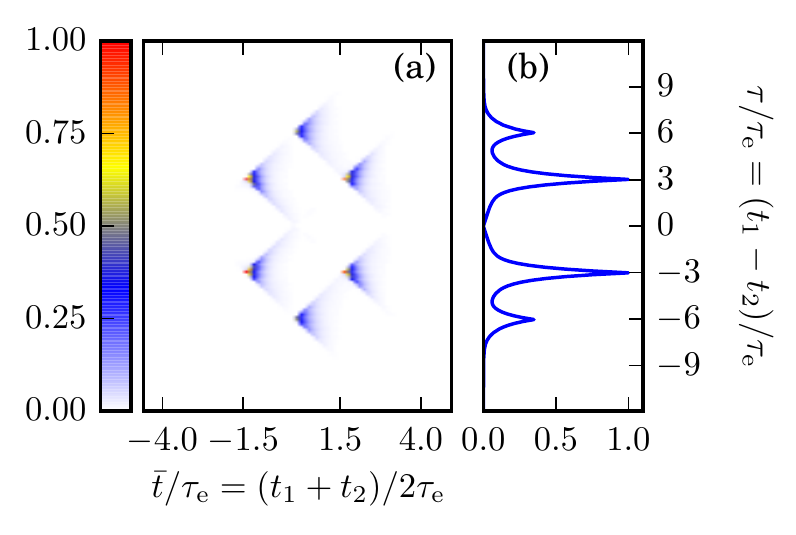}
\caption{\label{fig:correlations} 
(Color online) (a) Normalized density plot of $\Delta\Gee_S(t_1, t_2 | t_1, t_2)$ as a function of
$\bar{t} = (t_1+t_2)/2$ and~$\tau = t_1-t_2$ for
three Landau excitations
of duration $\tau_\text{e}$ and energy $\omega_{\text{e}}=3/\tau_{\text{e}}$
separated  by $\Delta t = 3\,\tau_\text{e}$. 
(b) Integral $\int \DGee_S \dd\bar{t} $ \textit{i.e.} average
correlations as a function of~$\tau$ only. Vanishing at $\tau=0$ is a
manifestation of Pauli principle.}
\end{figure}

As an example, 
Fig.~\ref{fig:correlations} presents 
$\Delta\Gee_S(t_1 , t_2 | t_1, t_2)$ 
for an ideal source emitting a Slater determinant built from
three Landau excitations~\cite{Ferraro:2014-2} of duration $\tau_\text{e}$ and energy
$\hbar\omega_{\text{e}}$ \footnote{Landau excitations are described by a truncated
Lorentzian in energy space: 
$\widetilde{\varphi}(\omega) = 
	\mathcal{N}^{-1/2} \,
	 \Theta(\omega)
	/(\omega - \omega_\text{e} + \frac{\ii}{2 \tau_\text{e}} )$ where
$\mathcal{N}$ is the normalisation constant.}
separated by $\Delta t = 3\,\tau_\text{e}$.
Correlations are maximal at the times associated with two
electron emissions and vanish around the diagonal $t_1 \simeq t_2$
as expected from Pauli principle.

\medskip

In the context of electron quantum optics, two-electron properties can be 
accessed through current noise measurement. But the
current noise~\cite{Mahe:2010-1,Parmentier:2012-1} of a source
contains contributions from terms~\eqref{eq:g2:class} and~\eqref{eq:g2:exch}. 
However, the HBT setup~\cite{Hanbury:1956-2,Bocquillon:2012-1} provides a more
direct access to $\DGee_S$.
By partitioning
electron and hole excitations emitted by the source against a reference
Fermi sea at an electronic beam splitter, it
dispatches the incoming coherences into the two outgoing
channels numbered $1$ and $2$. Consequently,
the outgoing interchannel two-electron coherence 
is related to~$\DGee_S$ through
\begin{equation}
\label{eq:hbt:G2}
\DGee_{\text{out}}(1,t_1;2,t_2 | 1, t'_1 ; 2,t'_2)
	= R T \, \DGee_S(t_1, t_2 | t'_1, t'_2)
\end{equation}
where $R$ (resp. $T=1-R$) denote 
the reflexion (resp. transmission) probability of the QPC.
As expected from quantum optics, 
$\Delta S_{12}^{\mathrm{out}}(t,t')$ defined as the excess of current
correlation
\begin{equation}
S^{\mathrm{out}}_{12}(t,t')
 = \langle 
 i_{1}^{\mathrm{out}}(t) \, i_{2}^{\mathrm{out}}(t')\rangle
 -
 \langle i_{1}^{\mathrm{out}}(t) \rangle 
 \langle i_{2}^{\mathrm{out}}(t')\rangle\,
\end{equation}
is related to the diagonal part of $\DGee_S$ by
\begin{equation}
\label{eq:G2:hbt-correlations}
\Delta S_{12}^{\mathrm{out}}(t,t') =
(e \, v_F)^2 \, RT\, \DGee_S(t,t'|t,t')
-\langle i_1(t)\rangle\langle i_2(t')\rangle\,
\end{equation}
thus showing that correlation measurements in the HBT setup give a direct access
to the diagonal part of $\DGee_S$. 

\medskip

Let us now focus on the off-diagonal part of two-electron coherence
emitted by the source.
First of all, for a train of electronic wavepackets, Eq.~\eqref{eq:G2:train} shows
that the only off-diagonal contributions to $\DGee_S$ arise from terms of
the form
$\varphi_i(t_1) \varphi_j(t_2) \varphi_i(t'_1)  \varphi_j(t'_2)$
where $i\neq j$ and from their image under exchange operations
$(t_1 , t_2 ; t_1',t_2') \mapsto (t_2 , t_1 ; t_1', t_2')$ or
$(t_1 , t_2 ; t_1', t_2')\mapsto (t_2,t_1;t'_2,t'_1)$. 
Since the antisymmetry 
\begin{equation}
\label{eq:G2:antisymmetry}
\begin{aligned}
\DGee_S(t_1 , t_2 | t_1', t_2') &= 
	- \DGee_S(t_2, t_1 | t_1', t_2') \\
	&= -\DGee_S(t_1 , t_2 | t_2', t_1')
\end{aligned}
\end{equation}
is purely kinematical, 
the physical two-electron
coherence time corresponds to the decay of
$\varphi_i(t_1)\varphi_i(t'_1)\varphi_j(t_2)\varphi_j(t'_2)$
as a function of $\tau_1=t_1-t'_1$ and $\tau_2=t_2-t'_2$ and not from their
images under the above exchange operations.
Therefore, in this
particular situation, the two-electron coherence time is related to the durations
of the wavepackets that govern the single electron
coherence time scale.

On the other hand, quantum superposition of two time-shifted
Slater determinants introduces a new time scale for two-electron
coherence. The two-electron state 
$(\psi^\dagger[\varphi_a]\psi^\dagger[\varphi_b]|F\rangle+
\psi^\dagger[\varphi_c]\psi^\dagger[\varphi_d]|F\rangle)/\sqrt{2}$
with mutually orthogonal wavepackets leads to 
\begin{equation}
\label{eq:G2:pairs}
\DGee(t_1,t_2|t'_1,t'_2) = \frac{1}{2}\sum_{\substack{\text{pairs} \\
\{ij\}, \{kl\}}}
\Phi_{ij}(t_1,t_2)\Phi^*_{kl}(t'_1,t'_2) \,.
\end{equation}
Terms involving identical pairs correspond to
Eq.~\eqref{eq:G2:train} and their contributions are located close
to the diagonal.
Assuming
that $\varphi_{c,d}$ are the time-shifted by 
$\Delta T_{\text{tb}}$ of $\varphi_{a,b}$, terms with $\{i,j\}\neq \{k,l\}$ contribute to 
off-diagonal two-electron coherences, now
extending over~$\Delta T_\text{tb}$. These coherences are a signature
of the time-bin entanglement~\cite{Brendel:1999-1,Marcikic:2002-1} of the quantum superposition of 
$\psi^\dagger[\varphi_a]\psi^\dagger[\varphi_b]|F\rangle$ and
$\psi^\dagger[\varphi_c]\psi^\dagger[\varphi_d]|F\rangle$.

\medskip

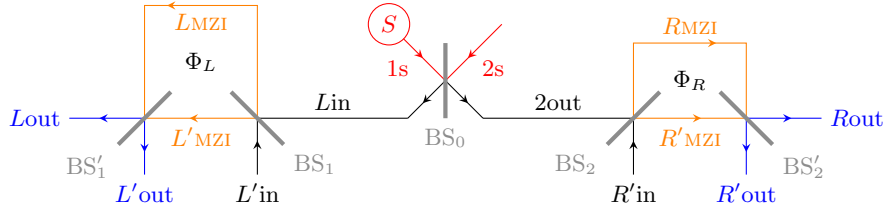
\begin{figure*}
\centering\small
\begin{tikzpicture}[scale=0.5,
  decoration={
    markings,
    mark=at position .6 with {\arrow{stealth}}}
  ]
\coordinate (O) at (0,0);
\coordinate (entree_MZI1) at (-5,-1);
\coordinate (sortie_MZI1) at (-8,-1);
\coordinate (entree_MZI2) at (5,-1);
\coordinate (sortie_MZI2) at (8,-1);

\coordinate (source1) at (-1.5,1.5);
\coordinate (source2) at (1.5,1.5);

\draw [red, postaction={decorate}] (source2) -- (O) 
  node [midway, below right] {2s};
\draw [red, postaction={decorate}] (source1) -- (O)
  node [midway, below left] {1s};

\filldraw[fill=white, draw=red] (source1) circle (.55);
\draw [red] (source1) node {$S$};


\draw [postaction={decorate}] (O) -- ++(-1,-1);
\draw (-1,-1) -- (entree_MZI1)
  node [midway, above] {$L$in};
\draw [postaction={decorate}] (O) -- ++(1,-1);
\draw (1,-1) -- (entree_MZI2)
  node [midway, above] {2out};
\draw [postaction={decorate}]  (-5,-2.5) node [below] {$L'$in}
  -- (entree_MZI1) ;
\draw [postaction={decorate}]  (5,-2.5) node [below] {$R'$in}
  -- (entree_MZI2) ;

\draw [orange, postaction={decorate}] (entree_MZI1) -- (sortie_MZI1)
  node [midway, below] {$L'$\textsc{mzi}};
\draw [orange, postaction={decorate}] (entree_MZI1) -- ++(0,3) -- ++(-3,0)
  node [midway, below] {$L$\textsc{mzi}}
  -- (sortie_MZI1);
\draw [orange, postaction={decorate}] (entree_MZI2) -- (sortie_MZI2)
  node [midway, below] {$R'$\textsc{mzi}};
\draw [orange, postaction={decorate}] (entree_MZI2) -- ++(0,2) -- ++(3,0)
  node [midway, above] {$R$\textsc{mzi}}
  -- (sortie_MZI2);

\draw (entree_MZI1) ++(-1.5,1.5) node {$\Phi_L$};
\draw (entree_MZI2) ++(1.5,1) node {$\Phi_R$};

\draw [blue, postaction={decorate}] (sortie_MZI1) -- ++(-2,0)
  node [left] {$L$out};
\draw [blue, postaction={decorate}] (sortie_MZI1) -- ++(0,-1.5)
  node [below] {$L'$out};
\draw [blue, postaction={decorate}] (sortie_MZI2) -- ++(2,0)
  node [right] {$R$out};
\draw [blue, postaction={decorate}] (sortie_MZI2) -- ++(0,-1.5)
  node [below] {$R'$out};

\draw [ultra thick, gray!90] (O) ++(0,1) -- ++(0,-2) node [below] {BS$_0$};
\draw [ultra thick, gray!90] (entree_MZI1) ++(-.7,.7) -- ++(1.4,-1.4) 
  node [below right] {BS$_1$};
\draw [ultra thick, gray!90] (entree_MZI2) ++(.7,.7) -- ++(-1.4,-1.4)
  node [below left] {BS$_2$};
\draw [ultra thick, gray!90] (sortie_MZI2) ++(-.7,.7) -- ++(1.4,-1.4)
  node [below right] {BS$'_2$};
\draw [ultra thick, gray!90] (sortie_MZI1) ++(.7,.7) -- ++(-1.4,-1.4)
  node [below left] {BS$'_1$};
\end{tikzpicture}
\caption{(Color online) Franson interferometer setup used to access two-particle coherence 
in the time domain. 
The left (L) and right (R) MZI are pierced by magnetic fluxes
$\Phi_{L,R}$ and have time delays $\delta t_{L,R}$. 
An electron source
is connected to the channel 1s whereas channels 2s, $L'$in and $R'$in are grounded.
The quantity of interest is the current correlation $\langle
i_{L\mathrm{out}}(t)\,i_{R\mathrm{out}}(t')\rangle$.
\label{fig:Franson}}
\end{figure*}

To be captured, 
quantum correlations contained in 
the off-diagonal parts ($t_1\neq t'_1$ and $t_2\neq t'_2$) must be
converted into measurable quantities. 
As discussed by Haack {\it et al.}~\cite{Haack:2011-1}, an ideal Mach--Zehnder interferometer (MZI)
converts off-diagonal coherence in the time domain into an average
current, \textit{i.e.} diagonal coherence.
This idea naturally leads to a Franson interferometer like setup~\cite{Franson:1989-1}
depicted on Fig.~\ref{fig:Franson}, in which MZI 
are added in each outgoing arm of a beam splitter.
Such a setup was introduced in electron quantum optics 
to evidence two-particle Aharonov--Bohm effect and 
entanglement generation~\cite{Splettstoesser:2009-1}.
Here, we use it differently, namely as a measurement device, as originally
proposed in photon quantum optics~\cite{Franson:1989-1,Brendel:1999-1,Marcikic:2002-1}. 

The current correlation between the two outgoing arms~$L$out and~$R$out picks up 
a magnetic flux dependence through the Aharonov--Bohm magnetic phases~$\Phi_{L,R}$.
The contributions depending on both magnetic fluxes, called twice Aharonov--Bohm terms,
involve delocalization on both MZI and
are the only one involving off-diagonal two-electron coherences in the time domain.
They give access to the intrinsic two-electron coherence of the source since 
\begin{widetext}
\begin{align}
\langle i_L(t_1)\,i_R(t_2)\rangle_{\text{2AB}}
& = (ev_F)^2 \, \kappa
\left(
\DGee_S(t_1-\delta t_L,t_2-\delta t_R 
|t_1,t_2) 
\ee^{\ii(\Phi_L+\Phi_R)}
+ 
\DGee_S(t_1,t_2 
|t_1-\delta t_L,t_2-\delta t_R) 
\ee^{-\ii(\Phi_L+\Phi_R)}\right.\nonumber\\
& + \left.
\DGee_S(t_1-\delta t_L, t_2 |t_1,t_2-\delta t_R) 
\ee^{\ii(\Phi_L-\Phi_R)}
 + 
\DGee_S(t_1,t_2-\delta t_R 
 |t_1-\delta t_L, t_2) 
\ee^{-\ii(\Phi_L-\Phi_R)}
\right)
\label{eq:G2outFranson:qtm} 
\end{align}
\end{widetext}
where $\delta t_{L,R}$ are the time delays within the two MZI and
$\kappa= R_0 \, T_0 \,R_1 \, T_1 \, R_2 \, T_2$ denotes the product of all the reflexion and 
transmission probabilities of the beam splitters.
Each contribution to the r.h.s. of Eq.~\eqref{eq:G2outFranson:qtm} 
can be isolated by a Fourier transform of current correlation with respect
to~$\Phi_{L,R}$. Low frequency correlations thus gives access to 
the real and imaginary parts of the integral over $(t_1,t_2)$ of 
$\DGee_S(t_1-\delta t_R, t_2-\delta t_L | t_1, t_2)$.

\begin{figure}
\includegraphics[scale=1]{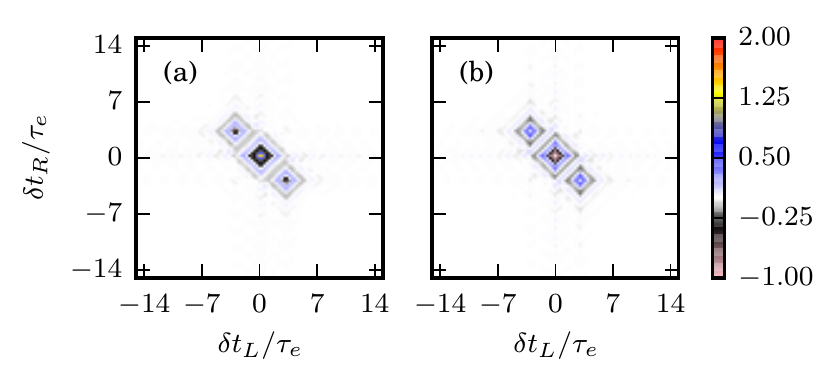}
\includegraphics[scale=1]{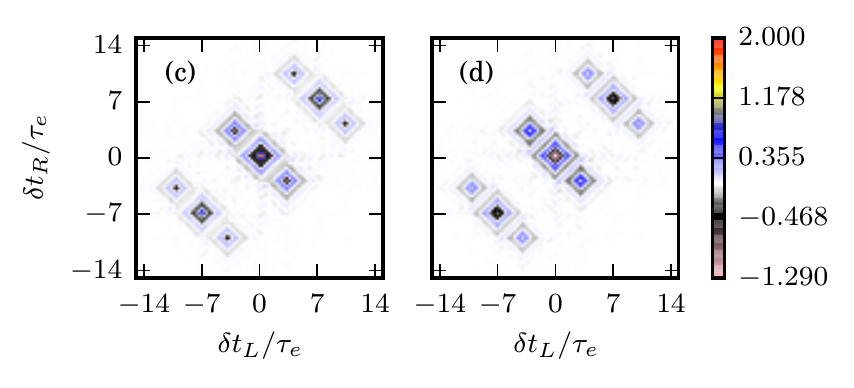}
\caption{\label{fig:output_franson}	
(Color online) Density plots for real (panels a and c) and imaginary (panels b and d) parts of the averaged excess two-electron coherence  
as a function of $(\delta t_L,\delta t_R)$,
extracted from the $\Phi_L+\Phi_R$ dependance of the
low frequency correlation~\eqref{eq:G2outFranson:qtm}. 
Panels (a) and (b): signal for a pair of Landau excitations 
of duration $\tau_\text{e}$ and energy $\omega_{\text{e}}=3/\tau_{\text{e}}$ 
separated  by~$\Delta t = 3\,\tau_\text{e}$. The plot is identical for 
a statistical mixture of two time-shifted identical electron pairs.
Panels (c) and (d): signal expected for a quantum 
superposition of two such pairs in two time-bins, separated
by~$\Delta T_\text{tb} = 7\tau_\text{e}$.
} 
\end{figure}

Fig.~\ref{fig:output_franson}a-b present these quantities
as functions of $\delta t_L$ and
$\delta t_R$ when $(S)$ emits 
a pair of Landau excitations separated by $\Delta t = 3\,\tau_\text{e}$.
Both plots show a central peak at $\delta t_L=\delta
t_R\simeq 0$ and satellite peaks at 
$\delta t_L \simeq -\delta t_R \simeq \pm 3\,\tau_{\text{e}}$. 
The central peak arises from single electron interferences which
are visible as soon as $\delta t_L$ and $\delta t_R$ are shorter
than the single electron coherence time.
The satellite peaks are due to non-local two-electron interferences 
between the two time-shifted electrons of the pair.
Their symmetry comes from the generic antisymmetry of 
two-electron coherence, Eq.~\eqref{eq:G2:antisymmetry}.
They spread along the line $\delta t_L = -\delta
t_R$ over the total duration of the train corresponding to
a classical correlation time and, since they do not depend on the pair's
emission time, they are the same for a statistical mixture of two time
shifted identical electron pairs.
Note that the noise signal is of the order of
the HOM signal for an HOM experiment $e^2f$ multiplied by $1/16$ due to the two MZI.
This is of the order of $3\times 10^{-30}~\mathrm{A}^2/\mathrm{Hz}$
for a 2~GHz source, and therefore within reach of state of the art current
noise measurements~\cite{Bocquillon:2013-1,Jullien:2014-1}.

For a superposition of time-shifted electron pairs, off-diagonal
two-electron coherences appear in the expected signals~\cite{Altimiras:2010-1} as shown
Fig.~\ref{fig:output_franson}c-d. 
In this case, the two-electron coherence time $\Delta T_{\text{tb}}$
corresponds to the spreading along the $\delta t_L = \delta t_R$ diagonal.
Considering interactions in the regime where electron/hole
excitations generated by the incoming electron pair are well separated
from the injected electronic excitations~\cite{Degio:2009-1} shows that,
for wavepackets of very short duration the
weight of $\delta t_L=\delta t_R$ satellite peaks is decreased by an
effective two electron decoherence arising from the overlap
of the imprints left in the effective environnement by the two
time-shifted electron pairs~\cite{Cabart:2016-1}.  The weight reduction of the $\delta
t_L=\delta t_R$ satellite peaks with respect to the central peak is thus
a direct measurement of two-electron decoherence~\footnote{In practice, a small
$\Delta T$ would be required for observing these peaks and a more
quantitative analysis required.}.

\medskip

To conclude, we have defined the intrinsic
two-electron coherence emitted by an electronic source, related it to 
two-electron wavefunctions. We have shown how to access its
off-diagonal part through current correlations in an ideal Franson
interferometer and how it gives access to two-electron decoherence.
The MZI being sensitive to decoherence,
this protocol suffers 
from the same limitations as the proposal 
by Haack {\it et
al.}~\cite{Haack:2011-1} for single electron coherence but this problem
can be circumvented by
replacing MZI by
devices converting off-diagonal single electron coherences into 
measurable quantities. For example, two-electron correlations in energy
could be accessed using energy filters~\cite{Altimiras:2010-1},
as put forward by Moskalets~\cite{Moskalets:2014-1}. Such measurements
would also probe two electron relaxation and
decoherence~\cite{Cabart:2016-1}.
Correlations of finite frequency currents seem promising 
since interaction effects in the detection stage
are encoded in finite frequency admittances~\cite{Bocquillon:2013-2}.
Finally, in the same way Hanbury
Brown and Twiss bypassed the challenges
posed by amplitude interferomery by moving to
intensity interferometry~\cite{Hanbury:1956-1},
two-electron coherences could be accessed
through an appropriate measurement of correlations of current noise.
This extends the idea of single electron tomography~\cite{Degio:2010-4} to
two-electron coherence.
Measuring current noise correlations in mesoscopic 
conductors is notoriously difficult~\cite{Zakka:2010-1} but we think that this method
deserves further investigation.

\begin{acknowledgments}
We thank M.~Moskalets and Ch.~Grenier
for useful discussions.
This work was supported by the ANR grants ``1shot'' (ANR-2010-BLANC-0412)
and ``1shot reloaded'' (ANR-14-CE32-0017).
\end{acknowledgments}


\end{document}